\documentclass[sigplan]{acmart}
\renewcommand\footnotetextcopyrightpermission[1]{}
\AtBeginDocument{%
  \providecommand\BibTeX{{%
    \normalfont B\kern-0.5em{\scshape i\kern-0.25em b}\kern-0.8em\TeX}}}

\setcopyright{acmcopyright}
\copyrightyear{2018}
\acmYear{2018}
\acmDOI{XXXXXXX.XXXXXXX}

\acmConference[Conference acronym 'XX]{Make sure to enter the correct
  conference title from your rights confirmation emai}{June 03--05,
  2018}{Woodstock, NY}

\DeclareFontFamily{T1}{calligra}{}
\DeclareFontShape{T1}{calligra}{m}{n}{<->s*[1.44]callig15}{}
\DeclareMathAlphabet\mathcalligra   {T1}{calligra} {m} {n}
\DeclareMathAlphabet\mathzapf       {T1}{pzc} {mb} {it}
\DeclareMathAlphabet\mathchorus     {T1}{qzc} {m} {n}
\DeclareMathAlphabet\mathrsfso      {U}{rsfso}{m}{n}

\usepackage{algorithm}
\usepackage[noend]{algpseudocode}
\usepackage{cancel}
\usepackage{graphicx}
\usepackage{subcaption}
\usepackage{tabularx}
\usepackage{tikz}
\usepackage[utf8]{inputenc}
\usepackage[english]{babel}
\usepackage{csquotes}
\usepackage{booktabs}
\usepackage{flushend}
\usepackage{stfloats}

\newcommand{\todo}[1]{\textcolor{gray}{\textbf{[TODO: #1]}}}

\newcommand{\jd}[1]{\textcolor{blue}{\textbf{[JD]: #1}}}

\algblockdefx{On}{EndOn}[1]{\textbf{on} $\mathtt{Receive}$ #1}{}

\newcommand*\circled[1]{\tikz[baseline=(char.base)]{\node[shape=circle,draw,inner sep=1pt] (char) {#1};}}

\newtheorem*{summary*}{Summary}

\newcommand{\systemname}{LØ}

\begin{document}

\title{LØ: An Accountable Mempool for MEV Resistance}

\author{Bulat Nasrulin}
	\affiliation{%
		\institution{Delft University of Technology}
		\city{Delft}
		\country{The Netherlands}
	}
	\email{b.nasrulin@tudelft.nl}
	\author{Georgy Ishmaev}
		\affiliation{%
	\institution{Delft University of Technology}
	\city{Delft}
	\country{The Netherlands}
	}
	\email{g.ishmaev@tudelft.nl}
    \author{Jérémie Decouchant}
		\affiliation{%
	\institution{Delft University of Technology}
	\city{Delft}
	\country{The Netherlands}
	}
	\email{j.decouchant@tudelft.nl}
	\author{Johan Pouwelse}
	\affiliation{%
		\institution{Delft University of Technology}
		\city{Delft}
		\country{The Netherlands}
	}
	\email{peer2peer@gmail.com}

\begin{abstract}

Possible manipulation of user transactions by miners in a permissionless blockchain systems is a growing concern. This problem is a pervasive and systemic issue, known as Miner Extractable Value (MEV), incurs highs costs on users of decentralised applications. Furthermore, transaction manipulations create other issues in blockchain systems such as congestion, higher fees, and system instability. Detecting transaction manipulations is difficult, even though it is known that they originate from 
the pre-consensus phase of transaction selection for a block building, at the \textit{base layer} of blockchain protocols. 
In this paper we summarize known transaction manipulation attacks. 
We then present \systemname{}, an accountable base layer protocol specifically designed to detect and mitigate transaction manipulations. \systemname{} is built around accurate detection of transaction manipulations and assignment of blame at the granularity of a single mining node. \systemname{} forces miners to log all the transactions they receive into a secure mempool data structure and to process them in a verifiable manner. Overall, \systemname{} quickly and efficiently detects reordering, injection or censorship attempts. Our performance evaluation shows that \systemname{} is also practical and only introduces a marginal performance overhead. 

\end{abstract}

\maketitle

\thispagestyle{plain} 
\pagestyle{plain}

\section{Introduction}



Enabled by blockchain technologies, Decentralised Finance (DeFi) tools and mechanisms have generated a lot of interest as building blocks for novel digital markets, both in terms of practical applications amounting to over 80 billion USD total value locked at the moment of writing, and in terms of significant research interest~\cite{xu2023sok}. Furthermore, these tools enable monetization mechanisms for the new paradigm of  Web3 development, providing alternatives to monopolistic centralised digital platforms. Decentralised exchanges, lending markets, derivatives, and other products built on permissionless blockchains are just some examples of these novel financial applications. However, these developments, are undermined by unresolved issues of transaction manipulations, such as \emph{censorship, injection}, and \emph{re-ordering} of transactions, at the expense of application users 
at underlying layers of blockchain protocols. 

This problem led to the notion of Miner Extractable Value (MEV)\footnote{Sometimes also referred to as Blockchain Extractable Value, or Maximum Extractable Value.}, which refers to the maximum revenue a miner can 
obtain from benign or manipulative transaction selection for block production~\cite{daian_flash_2019,qin_quantifying_2022}. This problem is a pervasive and systemic issue at large scale as exemplified by the Ethereum blockchain, where MEV transaction manipulations have generated over 320 USD million of revenue for bots and miners~\cite{wahrstätter2023time}. Furthermore, over 90\% of the blocks produced on Ethereum contain MEV transactions~\cite{qin_quantifying_2022}. Such manipulations not only undermine users' trust, but also induce systemic issues like congestion, inflated fees, and system instability~\cite{mazorra_price_2022}. 

We argue that the root cause of MEV is a lack of accountability at the base layer of permissionless blockchain protocols, sometimes referred to as 'dark forest'~\cite{qin_quantifying_2022}. By base layer, we refer to the processing steps that happen before consensus has to be reached on a block, such as sharing pending transactions (recorded in the mempool) with other miners and assembling them into blocks. In contrast to what happens at the consensus layer, at the base layer miners are expected to act as trusted parties. As such, a miner that creates a new block can arbitrarily select the transactions from its mempool. In practice, miners can therefore arbitrarily censor, inject or reorder transactions~\cite{gervais_tampering_2015}.

While this problem has received certain attention in the context of MEV mitigation tools, 
there are no comprehensive solutions preventing these types of transaction manipulations~\cite{piet2022extracting}. Most of the proposed solutions in this category focus on the application layer and on consensus layer mitigation tools~\cite{yang2022sok}. 
Many of these tools do not prevent MEV attacks but rather aim to mitigate them. The most well known approach, Proposer Builder Separation (PBS)~\cite{flashbots}, 
is implemented with the Flashbots middleware on Ethereum and does not prevent MEV, but only the redistribution of its associated revenues. Some proposed theoretical solutions, such as fair ordering consensus protocols~\cite{kelkar2021themis}, prevent transaction manipulations. However, these algorithms assume permissioned settings and small network sizes, and require important modifications of the blockchain consensus layer.    

As transaction manipulations arise from the lack of accountability at the base layer of blockchain protocols, we argue that comprehensive mitigation of MEV requires addressing trust assumptions at this particular layer. To address them, we design \systemname{}, an \textit{accountable mempool} protocol.  


In \systemname{} miners become accountable for the process of transaction selection and ordering. As new transactions are propagated among miners, they exchange and record commitments on the content of their mempools with each other. New transactions are shared in bundles, and commitment is recorded on a whole transaction bundle. This provides a local partial ordering of transactions.
Our system is based on \emph{pairwise commitments} that are exchanged during a mempool reconciliation phase, which is executed before the consensus protocol. This allows miners to witness each others' transaction selection and commit to a particular order and set of transaction that they will use for block generation. Therefore, \systemname{} ensures that any transaction manipulation, such as transaction censorship, injection and reordering, can be detected and proven by a correct node. 

Our system is agnostic to a specific type of consensus protocol in a permissionless blockchain system. It can be seamlessly integrated with existing blockchain solutions, as a relatively simple modification of a Peer-to-Peer (P2P) protocols that propagates transactions and blocks.
%
%
%
%
In addition, it does not require any additional cryptographic setups, and it does not impose a significant performance overhead. 
We leverage Minisketch data structure for the reconciliation of mempools to implement bandwidth-efficient commitments ~\cite{naumenko2019bandwidth}.
\\

This paper makes the following contributions: 

    $\bullet$ We identify key types of transaction manipulation attack primitives at the base layer. We propose a new taxonomy based on these attack primitives that can grasp all potential MEV attacks. We discuss the stages of the transaction processing pipeline that allow for these manipulations by miners. (\S\ref{sec:txmanipulation}).
    
    $\bullet$ After describing our system model (\S\ref{sec:sysmodel}), we provide an overview of \systemname{}, an accountable base layer protocol specifically designed to prevent transaction manipulations. \systemname{} is built around accurate detection of transaction manipulations and assignment of blame at the granularity of a single mining node. We discuss specific policies targeted at the detection of different MEV manipulations in (\S\ref{sec:overview}).
    
    $\bullet$ We detail how \systemname{} detects transaction manipulation attacks and potential mechanisms for the enforcement of these policies (\S\ref{sec:dealing}). We further discuss possible attacks against accountability in \systemname{}.
    
    $\bullet$ We present our performance evaluation, which demonstrates that \systemname{} is practical. It is both bandwidth and memory efficient. For example, it only requires up to 10 MB of additional storage for a network of 10,000 nodes and a workload of 20 transactions per second. At the same time, it is at least four times more efficient than the classical flooding-based mempool exchanges (\S\ref{sec:eval}).

\section{Transaction Manipulations at the Base Layer}\label{sec:txmanipulation}



\begin{figure*}
    \centering
    \includegraphics[width=1\linewidth]{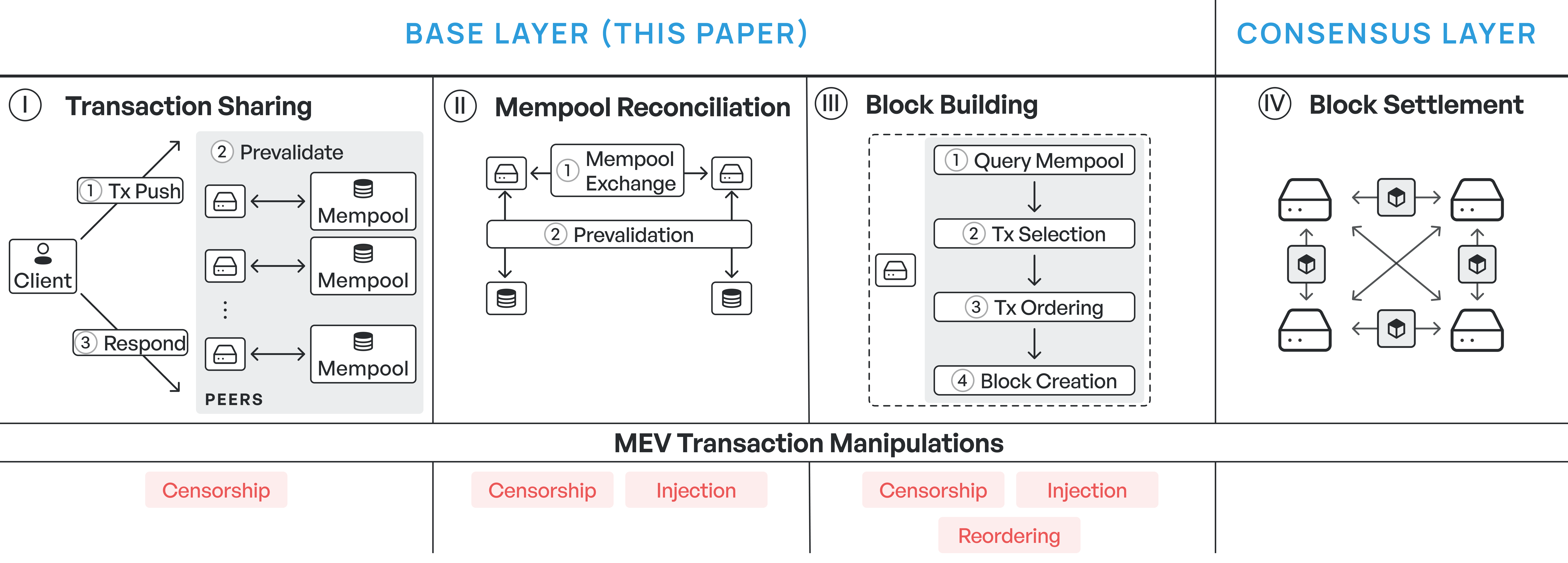}
    \caption{Functional modules of a permissionless Blockchain divided into two layers (base and consensus) and four stages.}
    \label{fig:blockchain_model}
\end{figure*}

We distinguish the \emph{base layer} of a blockchain system from its consensus layer. In the complete life-cycle of a transaction from its creation to its inclusion in a blockchain, the base layer corresponds to the steps that precedes the block consensus phase as illustrated in Fig.~\ref{fig:blockchain_model}. These steps include the creation of the transaction and its initial sharing, its inclusion in the mempools, the reconciliation of the mempools between miners, and the inclusion of the transaction in a candidate block. 

We emphasize that the block-building phase, where a miner selects transactions that it includes in a candidate block, is a pre-consensus phase. Indeed, while sometimes block building is described as part of blockchain protocols, it is strictly speaking not a part of the consensus mechanism as blocks can be produced offline, as illustrated by PBS in Ethereum and selfish mining in Bitcoin~\cite{eyal2018majority}. We further distinguish the base layer from the network layer of blockchain protocols, as the latter is required in all transaction processing phases, including during consensus. 

The base layer typically provides much lower guarantees against misbehaving nodes than the consensus layer. Miners only conduct checks on the validity and priority of transactions (which is related to miners fee) and add it to a local pool of unconfirmed transactions referred to as the 'mempool'~\cite{Wang_ethna_2021}. However, miners are considered to be trusted parties with regard to the selection, withholding, and ordering of transactions~\cite{gervais_tampering_2015}. Therefore, all phases of the transaction life-cycle that precede consensus allow transaction manipulations. 


\subsection{Transaction Manipulation Primitives}

We consider practical attacks that include reordering of transactions by miners. These attacks  have been observed in practical settings and described in academic works that relate to  MEV~\cite{yang2022sok}. In practice, these attacks combine different types of transaction ordering manipulations. A common taxonomy of MEV attacks is application-specific and depends on the source of attack revenue. Well known attack types include \emph{sandwich-attacks, front running, back running}, which are associated with decentralized exchanges,  \emph{sniping}, which is associated with Non Fungible Token auctions, and \emph{liquidations}, which are associated with collateralized loan protocols. This taxonomy evolves as new MEV attacks rapidly emerge with new applications.

In this paper we consider a different taxonomy focusing on specific attack primitives on the base layer. These primitives allow a broad range of MEV, either on their own or in combination. Namely: \emph{censorship, injection}, and \emph{re-ordering} of transactions. 

\textbf{Censorship}. Censorship is the ability of a miner to delay or ignore new transactions. Censorship can enable different financially motivated MEV attacks, such as \emph{sniping}, executed alone or in combination with other primitives. For example, when receiving transactions for a bid in Non Fungible Token auctions, a faulty miner can censor competing transactions to become the auction winner. This censorship mechanism can take place either during the mempool inclusion phase, or during the block inclusion phase.

\textit{Mempool Censorship}. Faulty miners can ignore transactions received from some other nodes, and exclude their valid transactions from their mempool. We assume that a faulty miner either provides a fake transaction reception acknowledgment, or does not acknowledge it at all. This type of attack enables censorship at the level of a mempool~\cite{yang2022sok}, and  facilitates transaction manipulation based on front-running. 

\textit{Blockspace censorship}. Faulty miners can exclude valid transactions from blocks, even after acknowledging their reception and including it in the mempool. This enables transaction censorship at the level of blockspace. 

\textbf{Injections}. We assume that honest miners include transactions received from other nodes in new blocks in a deterministic order. Honest miners can also add their own new transactions, under the assumption that updated mempool commitment  is shared with other nodes and acknowledged. Faulty miners inject new transactions in blocks in an arbitrary manner, without prior sharing of the updated mempool and without acknowledgements. 
This type of attack can result in certain types of transaction manipulations such as \textit{front-running, sandwich, back-running}~\cite{daian_flash_2019}. 

\textbf{Reordering}. Faulty miners can reorder transactions in a mempool and in a block in a way that deviates from a protocol and violates expectations of other nodes. Reordering is different from an injection attack, since a faulty miner does not add new transactions itself, but manipulates the order of transaction received from other nodes.  

\subsection{Transaction Processing Stages}

Attacks can happen at different stages of the transaction life-cycle. We model the processing of a transaction in a generic blockchain system in Fig.~\ref{fig:blockchain_model}. This processing happens in four stages: (I) initial transaction sharing, (II) mempool reconciliation, (III) block building and (IV) block settlement. In the following we describe each stage, and discuss the corresponding attacks 
that enable transaction manipulations such as MEV.  

\textbf{Stage I. Initial transaction sharing.}
A transaction is first created at the client side. The client signs the transaction with its private key. The transaction contains all the required context to be processed by miners, such as signature, UTxO address, execution commands, transaction fee, etc. 
The client shares the transaction with a subset of peers that it personally knows or whose identity is publicly known (step \circled{1}). The peers receive the transaction and attempt to prevalidate it (step \circled{2}). Our system is agnostic in respect to the choice of specific consensus protocol which will define the requirements for transaction prevalidation. For example, successful prevalidation of a transaction may require: valid signature from a client, sufficient amount of in a client account, and inclusion of a sufficient transaction processing fee.

Miners that successfully prevalidate a transaction insert it in their local mempool storage. Optionally, miners might respond to the client with the transaction status, to acknowledge inclusion of a transaction in a mempool (step~\circled{3}). Also optionally, client can query a miner to get an acknowledging of transaction inclusion in a mempool. A malicious peer can censor the transaction at the point of prevalidation, without adding it to the mempool performing \textit{Exclusion From Mempool}. For example, a peer can exclude a client based on its id, e.g. all transaction originating from a specific address. At the same time, client and peer can collude to include an invalid transaction into a mempool.

\textbf{Stage II. Mempool reconciliation.}
\label{mempool_rec_sec}
At this stage peers share their transaction mempools (step~\circled{1}). 
Typically, a mempool exchange is implemented to first share the transaction ids, and only later selectively share the transaction content of the corresponding ids. Once a miner receives the transaction content, it prevalidates the transaction (step~\circled{2}), similarly to stage I.
In theory, this stage allows miner to converge to the same transaction set for any peer-to-peer network.   
Unfortunately, in practice, there is no guarantee that miners will converge.  Client can be partially or completely excluded from learning particular transactions when communicating with malicious peers. Moreover, miners can inconsistently exchange their mempools. Finally, without a requirement for the mempool reconciliation, malicious miner can exclude or include any transaction without being detected by other miners. Different types of \textit{Injection attacks} and \textit{Exclusion attacks} can be performed by faulty miners at that stage. For example, a malicious miner receiving a high-fee transaction can withhold it from sharing with other nodes in order to include it in own block later.

\textbf{Stage III. Block building.}
 Upon creating a block, a miner populates it based on information stored in its local mempool data (step~\circled{1}). For each block, the miner selects a subset of transactions to fill up the blockspace (step~\circled{2}). The selected transactions are included in the block in a specific order chosen by the miner (step~\circled{3}). A final block contains additional metadata, like signature, nonce, or timestamp (step~\circled{4}). 
Most of the reported MEV is happening at the stage of block building. Indeed, miners can freely select, exclude, or order transactions to maximize their profit, performing \textit{Order manipulation} and \textit{Blockspace censorship}. 

\textbf{Stage IV. Block settlement.}
\systemname{} is agnostic to the specific consensus process to finalize the blocks. We model miner selection as a random process, where a selected miner build its block and sends it to other miners. 
The attacks on this stage are extensively discussed in  previous works. The most discussed manipulations include block withholding, block reordering and equivocation attacks. We consider the accountability on this stage out of scope. Our solution can be combined with other solutions addressing the manipulations on this stage, such as Polygraph~\cite{civit2021polygraph}.

\section{System Model}
\label{sec:sysmodel}

This section describes our system model, which is the classical one for blockchain protocols .

The mining nodes (miners) belong in a set $\Pi = \{p_1,p_2, \dots\}$ and communicate with each other by exchanging messages over the network. We assume that each miner is equipped with a cryptographic key pair, and is uniquely identified by its public key. Nodes have access to a cryptographic signature scheme and messages are authenticated.

\noindent \textbf{Communication Overlay.} Nodes form an undirected communication graph that is not assumed to be fully connected. Nodes are free to unilaterally add or drop local connections. Nodes are able to leave and later rejoin the network. Nodes share messages to their overlay neighbors through their direct connections. We use notation $N_i$ to refer to the neighbors of a node $p_i$, i.e., the nodes that are currently directly connected with it.

\noindent \textbf{Bootstrap and Peer Discovery.}
We assume that nodes that join the system are able to contact bootstrap nodes that facilitate node discovery. When (re)joining the network, each correct node requests a set of known active nodes from the bootstrap nodes. The bootstrap nodes are correct, i.e., they serve all nodes and unbiasedly propose a node from a set of locally known. As a result the nodes operate in one network.

\noindent \textbf{Continuous sampling}. Correct nodes continuously sample the network through a discovery procedure. \systemname{} is build on top of Byzantine resilient uniform sampling algorithm~\cite{bortnikov2009brahms}. Malicious nodes can delay the discovery, however, it is guaranteed that correct node will eventually be able to communicate.   

\noindent \textbf{Types of Nodes.}
In different consensus protocols nodes participating in block creation can be called validators, proposers, builders, etc. Here we only consider the role of block creator and refer to the nodes that create blocks as miners. For the sake of simplicity we do not consider light clients, which our model can trivially cover without modifications.  

Miners can create new transactions, and they can also propose new blocks with ordered transaction to be included in the blockchain. All nodes maintain a list of unconfirmed transactions (mempool) and exchange it with other nodes in the network through messages.

\subsection{Attacker Model}

In our network, each node is either correct or faulty. Correct nodes adhere to the reference protocol without data tampering and generate valid messages. Faulty nodes, on the other hand, can deviate arbitrarily from the reference protocol.

We assume that a faulty miner can execute any of the transaction manipulations we previously described: censoring transactions, injecting new transactions out-of-order, or deviating from the canonical transaction order~\cite{zhou2022sok}. These attacks can be carried out by a faulty miner in a naive way by sending the same message (e.g., a reordered set of transactions) to all neighboring nodes, or they can attempt to evade detection of manipulations by \textit{equivocating}, i.e., sending conflicting messages to different nodes.

\subsection{Accountability}

We consider the standard accountability property for distributed systems and protocols~\cite{haeberlen2007peerreview}. We define accountability as the ability to detect transaction manipulations and assign blame at the granularity of a single mining node.

In asynchronous environments, an adversary can try to evade detection as it is challenging to distinguish between a misbehaving node that deliberately ignores requests and a slow node. To circumvent this difficulty, we divide blames into two types: \emph{suspicions} and \emph{exposures}. An exposure is a verifiable proof of misbehavior, while a suspicion is a lack of response to a request.

We consider two desirable properties of accountability:

\textbf{Accuracy}: (1) \textit{Temporal.} No correct node is perpetually suspected by a correct node, and (2) \textit{No false-positives.} No correct node is exposed as misbehaving by other nodes.

\textbf{Completeness}: (1) \textit{Suspicion completeness.} Every misbehaving node that ignores requests is perpetually suspected by all correct nodes. (2) \textit{Exposure completeness.} Given an exposure message on node $p_i$, every correct node exposes node $p_i$ as misbehaving.




\section{\systemname{}: Accountable Base Layer }~\label{sec:overview}

In this section we present~\systemname{} which achieves accountability at the base layer. Specifically, \systemname{} is implemented as a modification of mempool reconcilation and block building stages.  

\subsection{New Explicit Policies at the Base Layer}

\begin{table*}[htbp]
\centering
\begin{tabular}{|c|c|c|}
\hline
\textbf{Addressed Manipulation} & \textbf{Current implicit policies} & \textbf{New explicit policies} \\
\hline
Censorship &
Unreliable Transaction Gossip & Inclusion of All Transactions  \\
\hline
Injection &
Out-Of-Order  Transaction Selection & Transaction Selection  in Received Order  \\
\hline
Reordering &
Arbitrary Order in a Block & Verifiable Canonical Order in a Block  \\
\hline
\end{tabular}
\caption{Implicit policies in the base layer of typical permissonless blockchain and the new explicit policies we replace them with to detect transaction manipulations. }
\label{tab:policies}
\end{table*}

This section introduces \systemname{}, our accountable base layer protocol for permissionless blockchains. \systemname{} improves over the `vanilla' mempool reconciliation and block building protocols of permissionless blockchains (stages 2 and 3 of Fig.~\ref{fig:blockchain_model}).   

To enable accountability we require to modify some currently implicit or ill-defined polices at the base layer. Our observation is that current implementations of blockchain systems use implicit policies that significantly complicate the detection of transaction manipulations. First, a transaction censorship is not possible to attribute to a miner given an unreliable transaction relay. Every miner has its own relaying policy, and even perfectly correctly behaving nodes may choose not to relay anything at all. Second, miners can build a block with any transactions from the mempool, or even inject new transactions during the block creation. Third, there is no `canonical order' inside a block, allowing for any type of reordering.    

Instead of these ill-defined policies we propose three alternative explicit policies to enable the detection of any transaction manipulations, as presented in Table~\ref{tab:policies}. In a nutshell, \systemname{} introduces three new explicit policies: \emph{Inclusion of All Transactions}, \emph{Transaction Selection in Received Order}, and \emph{Verifiable Canonical Order in a Block}. Transaction manipulations are detected as violations of our explicit policies during the mempool reconciliation, or when inspecting the content of a block.  

\noindent \textbf{Inclusion of All Transactions}. Each miner includes all valid transactions it encountered during the system run in its locally maintained append-only transactions set. Once two nodes are connected they directly exchange their known transactions. The transaction exchange is implemented as a sequence of set reconcilations. The miners exchange multiple transactions in one transaction bundle. This allows two nodes to efficiently obtain the transactions they are missing and as a result end up with the same transaction sets. 

The key ability of \systemname{} is that after a successful round of reconciliation both correct nodes are ensured to have a common set of observed transactions. To ensure that none of the transactions is censored and all processed in the same way miners keep all valid transactions they encounter. Miners commit to be able to reveal all transactions they know about, if necessary.   

\noindent \textbf{Transaction Selection in Received Order}. During the reconciliation process, each miner commits on the order it received a transaction bundle from another miner. To mitigate any out-of-order injections, the miners are required to process the transactions following their insertion order in their mempool. As miners learn and commit on their mempool transactions, the transactions are then naturally ordered according to the order with which they were received.  
    
\noindent \textbf{Verifiable Canonical Order in a Block}. Transactions that are inserted into a newly created block are selected according to a deterministic process. In more details, committed transaction bundles are first assembled following sequential order. The order inside a bundle is then pseudo-random: transactions are shuffled using a known shuffling algorithm and an \emph{order seed} value. The order seed value is based on the hash of the last created block. 

\subsection{Mempool Reconciliation}
\label{sec:mempool_comm}

The mempool reconciliation process (cf. \S\ref{mempool_rec_sec}) forces miners to correctly share the transactions they accepted into their mempool.  In practice, \systemname{}'s mempool reconciliation uses two techniques: (i) anti-entropy gossip reconciliations~\cite{eppstein2011s, van2008efficient}; and (ii) signed commitments~\cite{arakala2007fuzzy, naumenko2019bandwidth}.

\begin{algorithm}
\caption{LØ on miner $p_i$.}\label{alg:reconciliation}
\begin{algorithmic}[1]
\State $\widehat{\mathzapf{C}_1},\ldots,\widehat{\mathzapf{C}_N} \gets \emptyset,\ldots,\emptyset$ \Comment{Last observed commitments}
\State $\mathzapf{E} \gets \emptyset$ \Comment{Set of exposed miners}
\State $\mathzapf{S} \gets \emptyset$ \Comment{Set of suspected miners}
\Procedure{$\mathtt{NeighborsSync}$}{}
\For{$p_j \in Neighbors(i)$}
\If{$\mathzapf{C}_i \setminus \widehat{\mathzapf{C}_j} \neq \emptyset$} \Comment{Peer j is outdated}
        \State $\mathzapf{S} \gets \mathzapf{S} \cup \{p_j\}$
        \State \emph{request} $\mathzapf{C}_j \supseteq \mathzapf{C}_i$ from $p_j$
\Else
\State $\mathzapf{S} \gets \mathzapf{S} \setminus \{p_j\}$
\EndIf
\EndFor
\EndProcedure
\On{$\mathzapf{C}_j$}{}
    \If{$\widehat{\mathzapf{C}_j} \subset \mathzapf{C}_j  $}
        \State $\widehat{\mathzapf{C}_j} \gets \mathzapf{C}_j$
        \State 	$\Delta \mathzapf{C}_{ji} \gets  \mathzapf{C}_j \setminus \mathzapf{C}_i$
        \If{$\Delta \mathzapf{C}_{ji} \neq \emptyset$}
        \State \emph{send} $H(C_j), C_i$ to $p_j$ \Comment{Commit}
        \Else
        \State \emph{send} $C_i$ to $p_j$ 
        \EndIf
    \EndIf
    \If{$\mathzapf{C}_j \setminus \widehat{\mathzapf{C}_j} \neq \emptyset \And \widehat{\mathzapf{C}_j} \setminus \mathzapf{C}_j  \neq \emptyset ) $} 
        \State $\mathzapf{E} \gets \mathzapf{E} \cup \{p_j\}$
        \State \emph{Broadcast} $\mathzapf{C}_j, \widehat{\mathzapf{C}_j}$   
    \EndIf
\EndOn
\On{$H(C_i), C_j$}{}
    \State \emph{send} $txs \in \Delta C_{ij}$ to $p_j$
    \State $\widehat{\mathzapf{C}_j} \gets \mathzapf{C}_j \cup C_i$
\EndOn

\end{algorithmic}
\end{algorithm}

\begin{figure*}[t]
\centering
\includegraphics[width=0.99\linewidth]{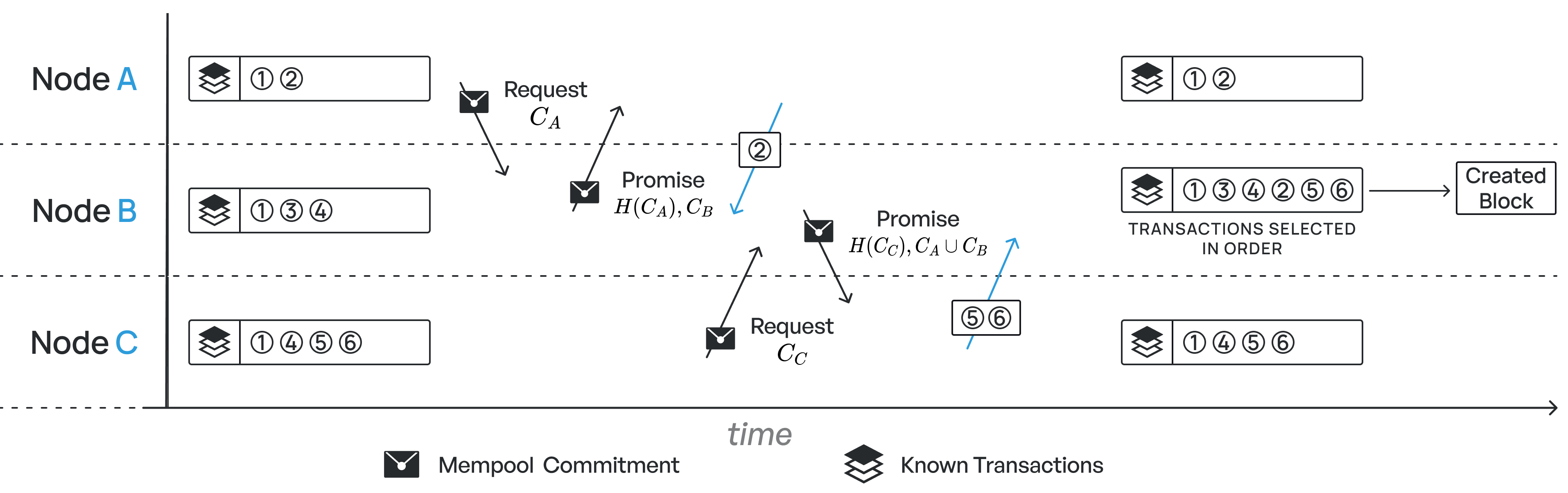} 
\caption{Example of a mempool reconciliation in \systemname{} between node $B$ and nodes $A$, $C$. Node $B$ preserves the transaction order for the block it eventually creates.   }
\label{fig:reconciliation}
\end{figure*}

Nodes maintain a mempool of all pending transactions and keep a record of all valid transactions they have ever received. Nodes reconcile their mempools to disseminate transactions throughout the system and generate commitments that are exchanged during mempool reconciliations. These commitments cover not only the transactions in the current mempool, but all valid transactions ever received by a node at the time of reconciliation.

Mempool reconciliation serves two purposes: (1) it allows miners to learn about new transactions from their neighbors; and (2) it ensures that miners commit to a specific transaction partial order during reconciliation. This partial order must be maintained during block creation. Miners mutually commit to the order by first exchanging a commitment. Miners are inherently motivated to receive transactions from other miners. However, they only disclose the transactions after their counterpart has committed to a specific order of transactions.

\textbf{Reconciliation Algorithm.} 
In Algorithm~\ref{alg:reconciliation} we provide \systemname{}'s pseudocode for a miner $p_i \in \Pi$. 

Periodically, miners require their neighbours to commit for new transactions  by sending them a request for a new commitment (line 4-10). While the request is pending, the node is suspected. We refer to $C_i$ as a commitment for the set of transactions included by miner $p_i$. At the same time, the commitment serves as a cryptographic checksum of included mempool transactions.  

During the reconciliation process, nodes first exchange a signed commitment $C$. After receiving the commitments of their neighbours, nodes calculate their transaction set differences with them (line 14).
Since the commitment is signed, it can later be used as a proof of inclusion of transactions---any receiver can use the commitment $C_j$ as verifiable evidence that node $p_j$ should have included transactions in its mempool. 

Our mempool reconciliation between a miner $p_i$ and miner $p_j$ works in two phases. In the first phase, miner $p_i$ sends to miner $p_j$ a request to commit to new set of transactions (line 8). A peer $p_j$ receives the request and responds either with its new $C_j$ that already includes all transactions  (line 18), or with a new commitment fixing locally the order of transactions $\Delta C_{ij}$, i.e., a promise to apply them immediately after all known local transactions $C_j$ (line 16). In the second phase, miner $p_i$ sends all the transactions corresponding to the $\Delta C_{ij}$ to peer $p_j$ (lines 23-25).  

All miners store at least the last received commitments from their overlay neighbors (line 13). On receiving a checksum ${C}$ it is first validated against previously received set $\widehat{{C}}$ (line 19-21). The set $\widehat{{C}}$ is grow-only and keeps all the transactions committed by the node. If ${C}$ is inconsistent against the previously reported messages $\widehat{{C}}$, the evidence of the faulty behavior is shared with other nodes (line 21). This inconsistency could happen for example when a faulty node is trying to hide a previously reported message or does not report a message received from other nodes. 

\textbf{Example.} Fig.~\ref{fig:reconciliation} illustrates a possible mempool reconciliation. 
Nodes $A$, $B$, and $C$ first exchange transaction commitments. Note that commitments can also be received indirectly, but this scenario is not included in Fig.~\ref{fig:reconciliation} for simplicity. Node $A$ sends a request, along with the mempool commitment $C_A$, to node $B$. Node $B$ reconciles commitment $C_A$ with its own $C_B$ and promises to include node $A$'s missing transactions immediately after all transactions $C_B$. Node $A$ promptly sends the missing transaction 2 to node $B$.
Shortly afterward, node $C$ reconciles with node $B$ in a similar manner. However, this time, node $B$ promises to include transactions of node $C$ only after the transactions 1,3,4,2. Let's assume that later, node $B$ creates a new block, possibly because it is elected as a consensus leader. Node $B$ must then select all transactions in the order of the commitment it made, which is 1,3,4,2,5,6.

\textbf{Implementation Details.}
\systemname{} employs \textit{Minisketch} and \textit{Bloom Clocks} to implement the mempool reconciliation protocol efficiently. A commitment in this context includes both the miner's Bloom Clock and Minisketch. These data structures serve two primary purposes: (1) they identify inconsistencies with the digests shared in previous rounds, and (2) they facilitate set reconciliation to identify a miner's unknown transactions.

A \textbf{Minisketch} is a data structure proposed for the bandwidth-optimized exchange of transaction sets between nodes in the Bitcoin network~\cite{naumenko2019bandwidth}. Initially proposed for the reconciliation of mempool data, it can also be used to optimize block propagation. In this protocol, a sketch serves as a "set checksum". The primary advantage of Minisketch is its ability to reconcile quickly and accurately. However, it has a downside: the requirement to decode the reconciled Minisketch, which can fail. In such cases, we repeat the process by dividing the set in half and sending two sketches.

A \textbf{Bloom Clock} is a space-efficient, probabilistic data structure used for the partial ordering of events in distributed systems~\cite{ramabaja2019bloomclock}. \systemname{} uses Bloom Clocks to swiftly detect inconsistencies between two sets. In rare cases, when a Bloom Clock fails to detect an inconsistency due to collisions, we resort to a hash checksum. We employ Bloom Clocks to speed up the verification of inconsistencies between two sets.

\begin{summary*}
The pairwise commitment scheme ensures that miners are committed to all transactions they discover according to the order with which they are received. 
\end{summary*}

\subsection{Block Building}\label{sec:block_build}

\begin{figure}
\centering
\includegraphics[width=1.0\linewidth]{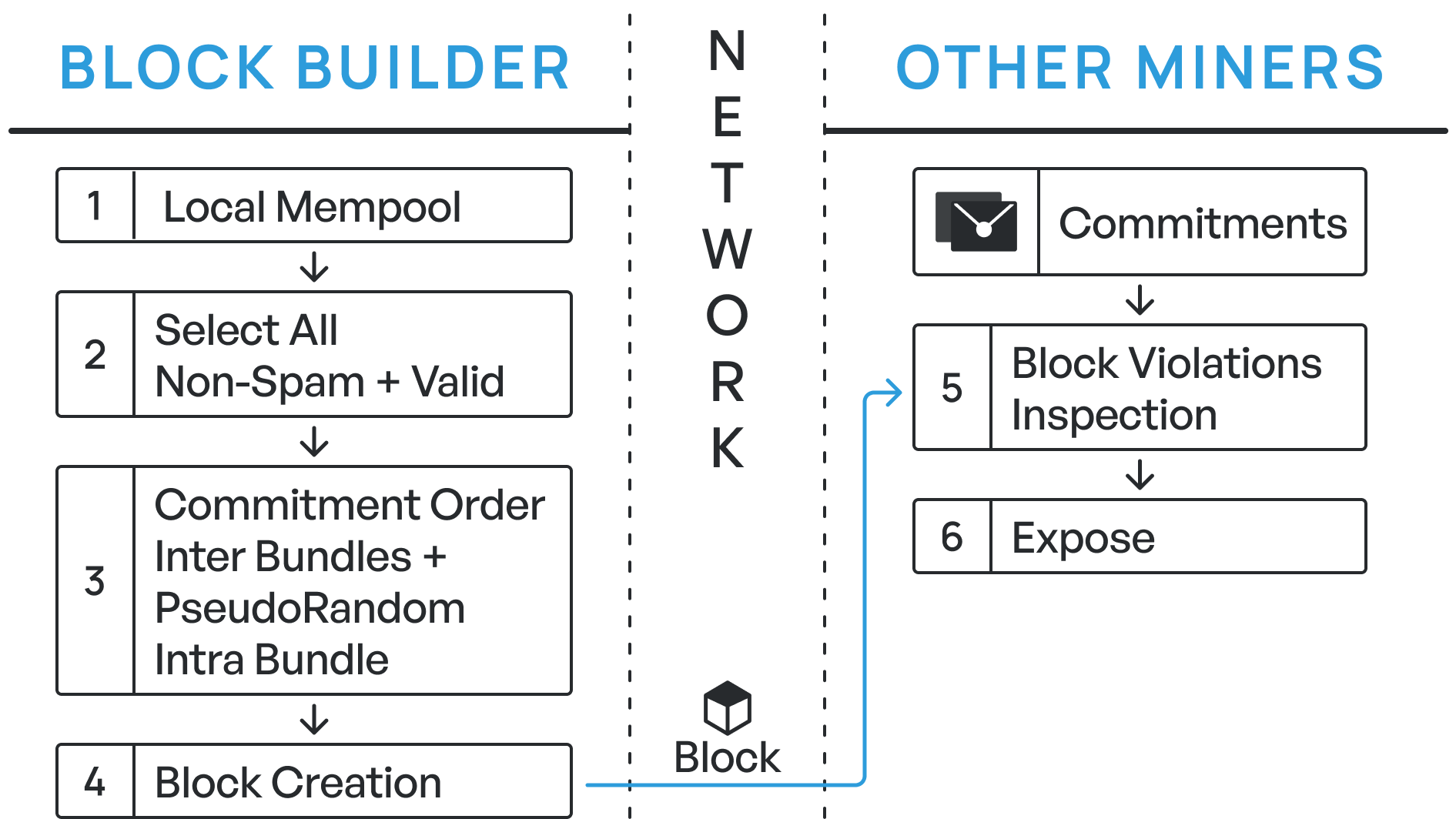} 
\caption{Block building and inspection in \systemname.} 
\label{fig:building}
\end{figure}

To avoid manipulations during the block building stage, we slightly modify the `vanilla' block building process with our new policies. The modified block-building process is shown in Fig.~\ref{fig:building}.

\textbf{Transaction Selection.} 
Peers select all transactions they encounter during the mempool reconcilation phase and that are included in the mempool (step 1). Miners must verify these transactions. The transactions that are not valid are not included in the block. The transactions that have fees lower than some threshold are not included in the block, and  are rejected (step 2).  

\textbf{Transaction Ordering.} The selected transactions are ordered in a verifiable canonical way (step 3).  Recall from the mempool reconciliation process that transactions are partially ordered with the commitments order as the commitments define the order between transaction bundles. We also define a deterministic pseudo-random  order function inside each of the bundle. We use a hash of previous block as a seed for the intra-bundle order function.  

\textbf{Block Inspection}. Next, a block is created (step 4) and shared with the network. 
Given the mempool commitments, any node can verify the produced block by inspecting its content with respect to the \systemname{} reference protocol (step 5). Note that block inspection is a separate process than block validation, and does not affect the block inclusion into the chain. Any violation exposes the block creator (step 6), by comparing the block content with the known commitments. Our protocol is agnostic to the specific punishment mechanisms, but we discuss some options in Section~\ref{sec:mitigation}. 

\begin{summary*}
During the block building process, miners select and order transactions deterministically. 
\end{summary*}

\section{Dealing with Attacks}
\label{sec:dealing}

This section presents an analysis of various attacks and discusses how the integration of detection mechanisms and a broad spectrum of enforcement tools can counter them.

\subsection{Detection of Transaction Manipulations}

Every node utilizes a block inspection module to detect violations. Nodes are required to disclose all their known transactions and they must consistently disclose each commitment or they run the risk of being identified as faulty. An inconsistency is detected when comparing two commitments, provided both sets contain at least one transaction.

Nodes are obligated to respond to commitment requests. Failure to do so results in an eventual fault suspicion by every correct miner in the network. Reconciliation messages and proposed blocks are validated against the protocol rules. Violations, such as censorship of particular transactions, commitment inconsistencies, or message tampering, can then be identified. Evidence of faulty behavior is disseminated across the network by correct miners. 

\textbf{Countering Attacks during Mempool Reconciliation.} Every node involved in a mempool reconciliation retains a signed commitment acquired from other nodes, which can be used to identify faulty nodes. Sufficient interaction with correct nodes in the network makes it virtually impossible for a node to manipulate its mempool and not be detected. The mempool reconciliation process thus ensures reliable detection of injection and mempool censorship attacks. A misbehaving miner attempting a front-running attack, for example, may inject a new transaction out-of-order. However, this attack is swiftly detected as the injected transaction would be inconsistent with previous commitments.

\textbf{Enhancing Detection Resilience.} After a mempool reconciliation between two miners, they can mutually detect each other's violations. Throughout the operation of the system, miners collect commitments from all their overlay neighbors. Consequently, an overlay neighbor can detect a violation. However, if an overlay neighbor is offline, it cannot broadcast the exposure message to other miners. To enhance resilience, miners share between each other a sample of the last commitments they received. This allows other non-neighbouring miners to also detect violations.

\textbf{Countering Attacks during Block Building.} The order function ensures that order manipulation attacks can be detected, as any block where the transaction order deviates from the canonical one will be detected. Similarly, a block-space censorship attack is detected as a deviation from the selection function rules.

\subsection{Suspicion and Misbehavior Sharing} 
\systemname{} provides guarantees that violation of block production rules can be reliably detected by other nodes in the block inspection process and that misbehaving node will be exposed. Our accountability mechanism provided in \systemname{} that consists of suspicions, equivocation detection, and exposure.

\begin{figure}
\centering
\includegraphics[width=\linewidth]{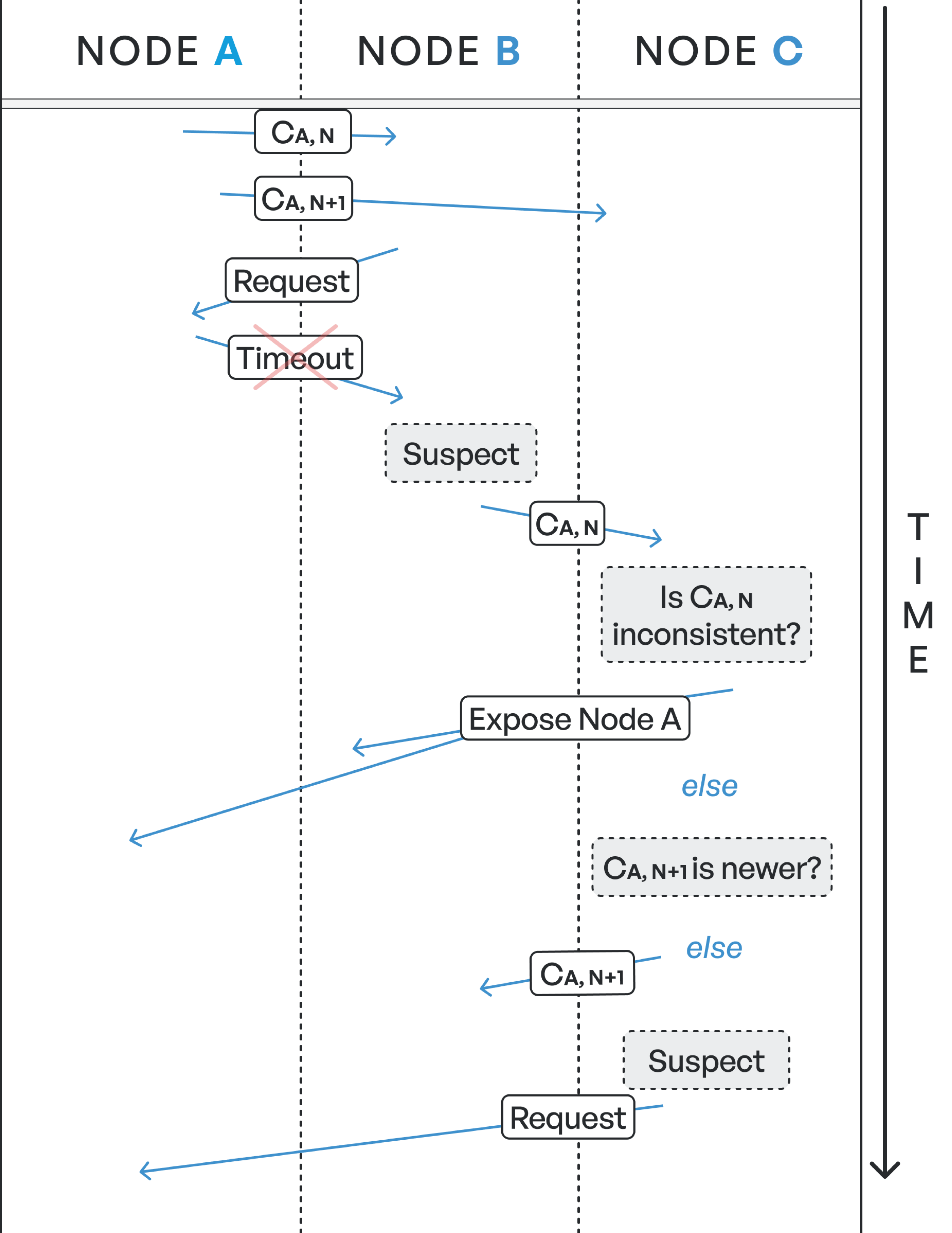}
\caption{Consistency check and suspicion mechanisms.}
\label{fig:inconsistency}
\end{figure}

\textbf{Suspicions.}
The \textit{Accountability Mechanism} incorporates liveness checks and propagates transaction commitments between nodes through indirect paths. If a node does not respond to transaction requests before a timeout, it is suspected by the requester. The requester may resend the request multiple times before suspecting the node. Correct nodes retain all pending requests. If a node is suspected, the requester broadcasts the suspected requestee's identity to other nodes, along with information on pending requests and the requestee's last known commitments.
A node may retrieve pending requests after a partition or a crash. Once it publicly responds to all pending requests, no correct node will suspect it.

In Fig. \ref{fig:inconsistency}, node $B$ has an earlier commitment $(C_{A, n})$ from node $A$. Node $C$ has the latest commitment $(C_{A, n+1})$ from node $A$. Node $B$ sends a request for a commitment on a particular transaction $\tau$ from node $A$, but does not receive a response. After a timeout, node $B$ suspects node $A$ and broadcasts a suspect status along with the latest commitment $(C_{A, n})$ from $A$ that is available to $B$ to its neighbors, in this case, to node $C$.

\textbf{Equivocation Detection.} A consistency check occurs when a node is suspected.
Commitments are append-only sets and thus follow chronological order. When a node has two commitments from a neighbor, it can easily detect any inconsistency between the previous commitment $n$ and the latest commitment $n+1$ using its bloom clock.
Nodes can receive commitments from other nodes both directly and indirectly. Consider an example of suspicion and consistency check in Fig.~\ref{fig:inconsistency}. Node $C$ receives two commitments originating from node $A$, i.e., commitment $(C_A, n+1)$ from node $B$, and $(C_A, n+1)$ from node $A$. Node $B$ has tried to get a commitment on transaction $\tau$ from A and suspects A because of the high response delay. Node $C$ will check whether $(C_A, n)$ and $(C_A, n+1)$ are consistent with each other.
\begin{itemize}
\item If these commitments are inconsistent, node $C$ exposes $A$ as a misbehaving node.
\item If $(C_A, n)$ and $(C_A, n+1)$ are consistent and $(C_A, n+1)$ already includes a commitment on a transaction $\tau$, then node $C$ will share the latest commitment $(C_A, n+1)$ with $B$.
\item If $(C_A, n)$ and $(C_A, n+1)$ are consistent but $(C_A, n+1)$ does not include a commitment on $\tau$, then $C$ will send a request for commitment on $\tau$ to $C$ and suspect $C$.
\end{itemize}

\begin{summary*}
Any mempool counterpart can submit a proof of misbehavior showing inconsistency between a mempool commitment and a produced block.
\end{summary*}

\subsection{Possible MEV Prevention Mechanisms}\label{sec:mitigation}

Reliable detection and blame assignment allow for MEV mitigation through the enforcement of policies. The choice of specific enforcement mechanisms depends on the consensus protocol. Given that \systemname{} is agnostic to the particular consensus algorithm used, a detailed analysis of specific enforcement mechanisms is beyond the scope of this paper.

For instance, in \textit{Proof-of-Stake} (PoS) consensus algorithms, various slashing strategies can be applied to misbehaving nodes~\cite{buterin2017casper}. Since validating nodes in PoS must invest a certain amount of funds to become validators, slashing of stake incurs a financial loss. For consensus algorithms based on the reputation of validating nodes, slashing of reputation can equivalently serve as a penalization mechanism~\cite{yu2019repucoin}. Misbehaving nodes can also be penalized at the network layer level, such as temporary disconnection from the network~\cite{gervais_tampering_2015}. In addition to penalizing misbehaving miners, detection allows the implementation of mechanisms for the rejection of blocks that deviate from the canonical transaction order~\cite{sheng2021bft}. However, this latter approach imposes significant trade-offs on the modification of the consensus protocol.

\subsection{Addressing Accountability Attacks}

In our model, we assume that miners are incentivized to learn about more transactions. This assumption aligns with empirical observations, as miner profitability correlates with their ability to discover new transactions~\cite{piet2022extracting}.

\begin{figure}
\centering
\includegraphics[width=0.99\linewidth]{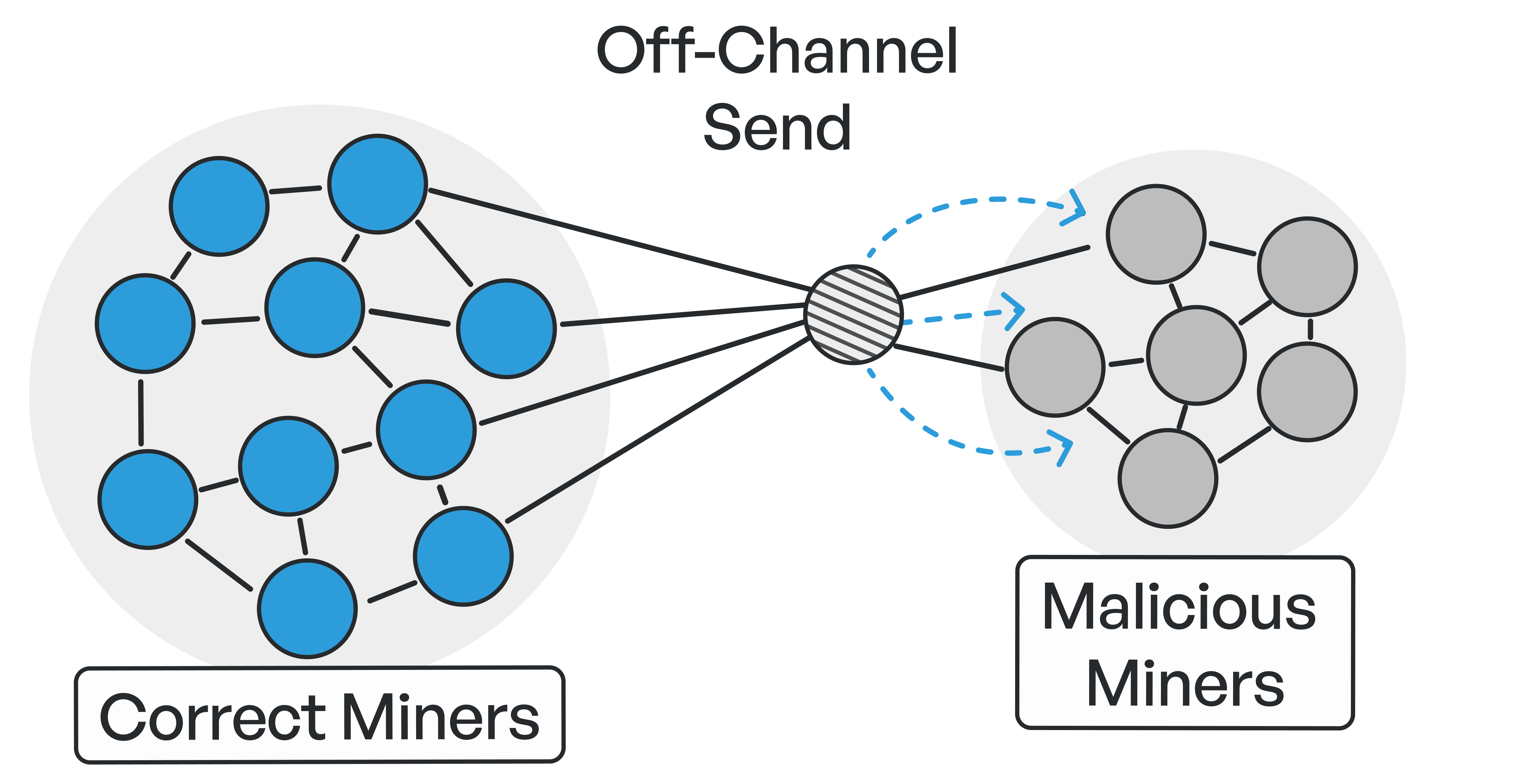}
\caption{Illustration of colluding malicious miners communicating off-channel to evade detection.}
\label{fig:collusion}
\end{figure}

In our system, miners only discover transaction content after exchanging commitments. Hence, by learning about new transactions, miners commit themselves to mempool commitments. However, a potential loophole exists for malicious miners.
A miner could conspire with an accomplice who does not interact with correct nodes to create a block for them using a manipulated transaction order. This attack is depicted in Fig.\ref{fig:collusion}.
A malicious miner can transfer a transaction, denoted as \emph{Tx}, to another colluding miner or to a Sybil miner under their control. Since a colluding miner has not exchanged commitments with the originator of \emph{Tx}, it can attempt to reorder or inject transactions, and propose an alternative block. However, this type of attack is impractical due to several reasons:
\begin{itemize}
\item Colluding miners can only front-run or back-run an entire original transaction bundle. Any attempt to inject, censor, or reorder transactions within a transaction bundle is eventually detectable by correct miners. This significantly restricts the attack granularity, a crucial factor for MEV profitability.
\item Colluding miners or Sybils cannot respond to queries from honest miners to evade commitments. They can only learn about new transactions via malicious nodes acting as a bridge. However, such a non-responding set of colluding miners is eventually detected and suspected.
\item Colluding miners or Sybil miners must have a high probability of becoming the consensus leader to include a specific transaction.
To increase this success rate, a substantial set of colluding miners or Sybils is required, which is costly considering the initial investment and the absence of profits from honest protocol participation.
\end{itemize}

Finally, to further mitigate the attack, one option is to require sufficient Proof-of-Interaction during block creation. Specifically, the block creator must also include signatures from a sufficient number of miners (based on mining power or stake), thereby proving recent interaction with them.   

\section{Evaluation}
\label{sec:eval}

This section presents our evaluation of \systemname{} focusing on its resilience against malicious nodes and the impact of such nodes on detection. We also discuss the overhead associated with \systemname{}.

\subsection{Experimental setup}

\systemname{} was evaluated experimentally on a national research cluster~\cite{bal2016medium}. Each server in the cluster is equipped with an Intel Xeon E5-2630 CPU with 24 physical cores operating at 2.4 GHz, hyper-threading enabled, and 128 GiB of main memory. The servers are interconnected via a Gigabit Ethernet network. \systemname{} was implemented in Python. We emulated realistic network latencies using netem\footnote{See \url{https://www.linux.org/docs/man8/tc-netem.html}} and incorporated ping statistics from 32 cities worldwide from the WonderNetwork dataset~\cite{WonderNetwork}. Each miner was assigned to a city in a round-robin manner.

Unless otherwise stated, the parameters for the reported experiment were set as follows: The experiment was conducted with $10,000$ nodes, generating a workload of 20 transactions per second, with each transaction being 250 bytes in size. The transactions were injected into our system based on a realistic dataset of Ethereum transactions~\cite{pierro2019influence}. Each experiment was repeated 10 times, and the average result of these runs is reported.

We constructed a connected topology where each node had eight outgoing connections and up to 125 incoming connections, in line with the default Bitcoin parameters. Every node attempted to reconcile with three random neighbors every second. The request timeout was set to 1 second. If a request was not fulfilled within this time, it was resent three times, after which the node was suspected of being faulty. The Minisketch size was set to 1,000 bytes, sufficient to reconcile a set difference of up to 100 transactions, allowing the Minisketch to fit into a single UDP packet. If reconciliation failed, all transactions were divided into two subsets, and the process was repeated with two sketches. The size of Bloom-Clocks was fixed at 32 cells (i.e., 68 bytes in total).

\subsection{Resilience}

\begin{figure}
\centering
\includegraphics[width=1.0\linewidth]{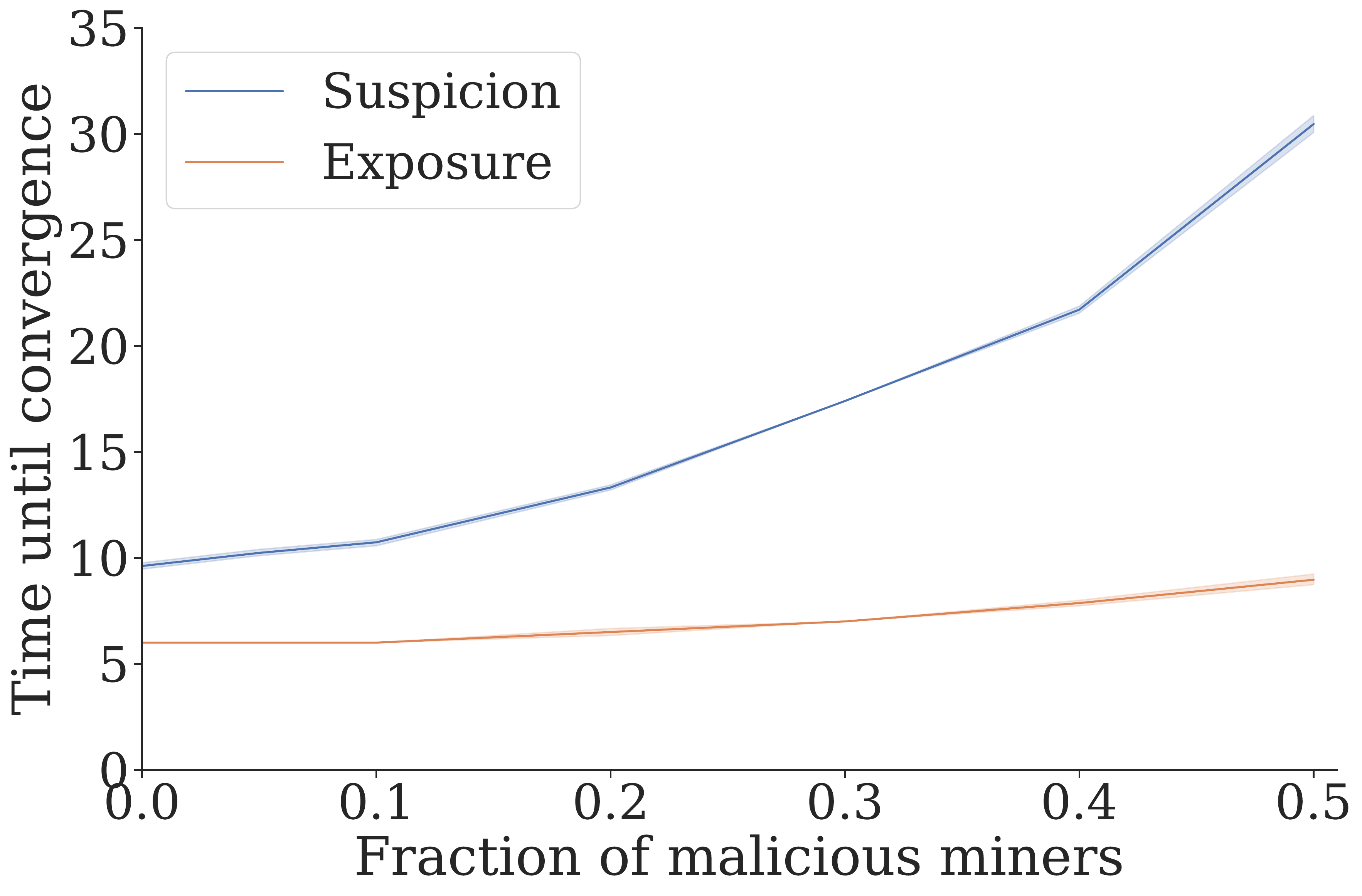} 
\caption{Time necessary to suspect or expose a faulty miner depending on the proportion of colluding censoring miners in the system.}
\label{fig:faultime}
\end{figure}

We assess the impact of colluding censoring miners on the network, specifically focusing on their effect on the convergence of correct nodes. In this scenario, malicious miners attempt to prevent correct nodes from learning about transactions, commitments, exposure, and suspicion messages. All malicious miners are assumed to be interconnected. For these experiments, we ensure that the correct nodes remain connected via some path in the network by initially running an unbiased sampling algorithm~\cite{bortnikov2009brahms,auvolat2021basalt}.

Fig.~\ref{fig:faultime} illustrates the time required for all correct nodes to converge, depending on the number of faulty nodes in the network. The presence of faulty nodes marginally increases the time needed for all correct nodes to learn about the exposure message, extending it to 6-7 seconds after the first miner detects and creates the message.

We also demonstrate how our system can detect faulty nodes that ignore requests. We report the time until every correct node suspects all faulty nodes (Fig.~\ref{fig:faultime}, `Suspicion'). As expected, the time until all faulty nodes are suspected is longer than the time required for nodes to discover an exposure message, as the nodes need to submit a request and wait for it to timeout.

\subsection{Transaction Latency}

\begin{figure}
\centering
\includegraphics[width=1.0\linewidth]{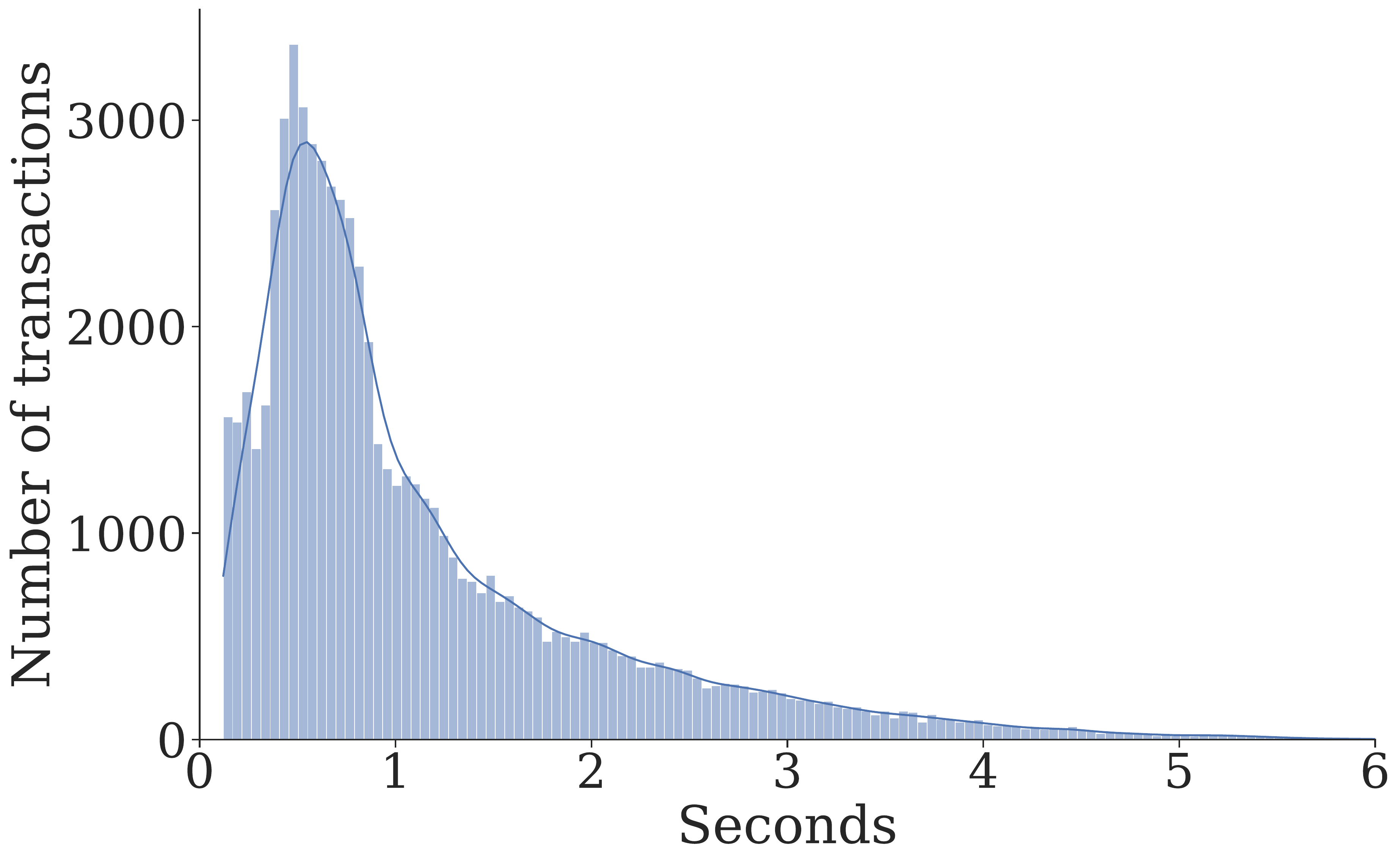}
\caption{Distribution of the time (s) required for a miner to insert a transaction into its mempool.}
\label{fig:time_mempool}
\end{figure}

We report the time necessary for miners to discover a transaction and include it in their mempool. The latency distribution is reported in Fig.~\ref{fig:time_mempool}. It appears that all nodes learn about the transaction after contacting 5 to 6 nodes. On average, a transaction is discovered by a node in 1.14\,s.

\begin{figure*}
    \centering
    \begin{subfigure}{.49\linewidth}
	    \centering
     \includegraphics[width=1.0\linewidth]{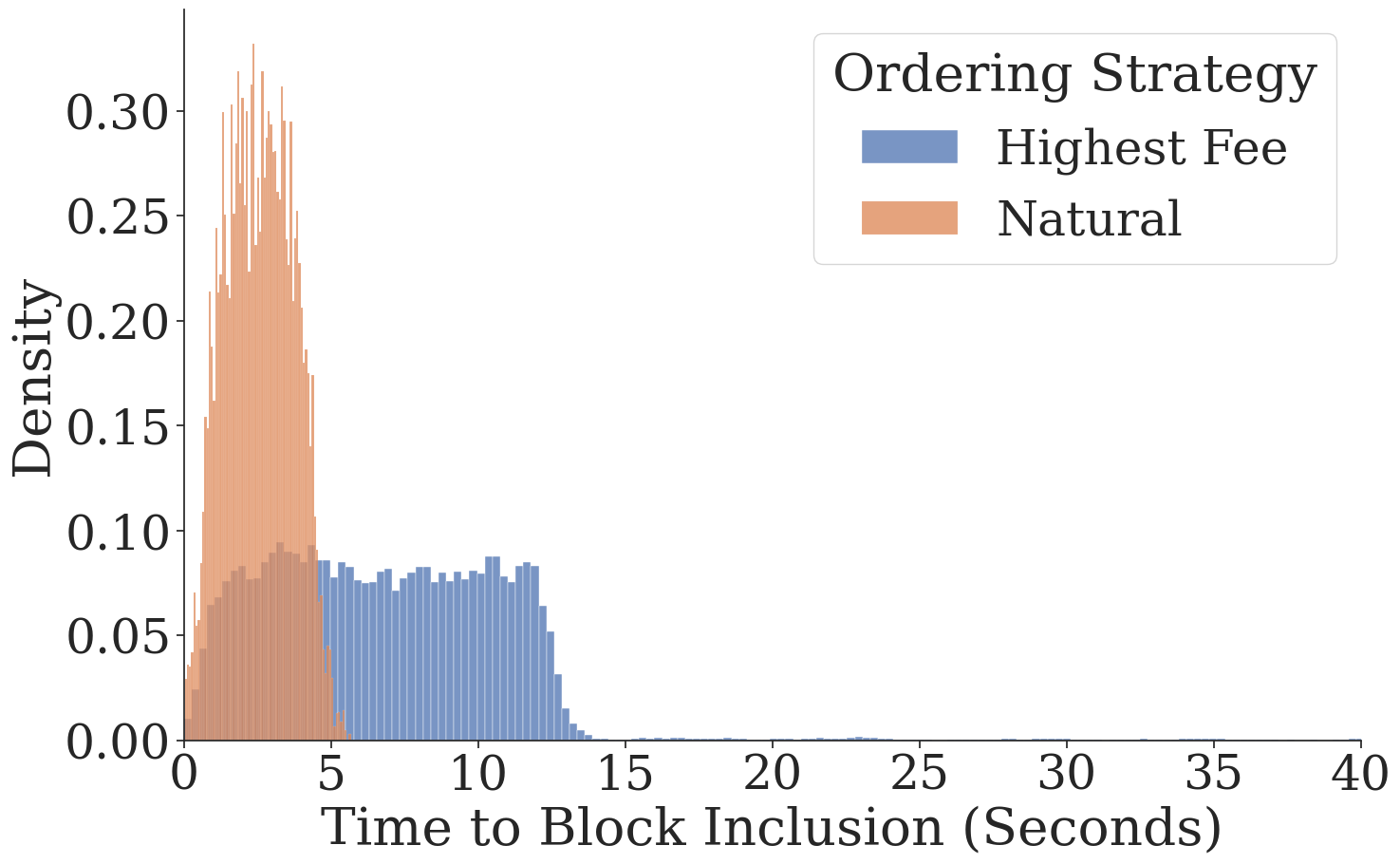}
	\end{subfigure}%
	\centering
	\begin{subfigure}{.49\linewidth}
	    \centering
    \includegraphics[width=1.0\linewidth]{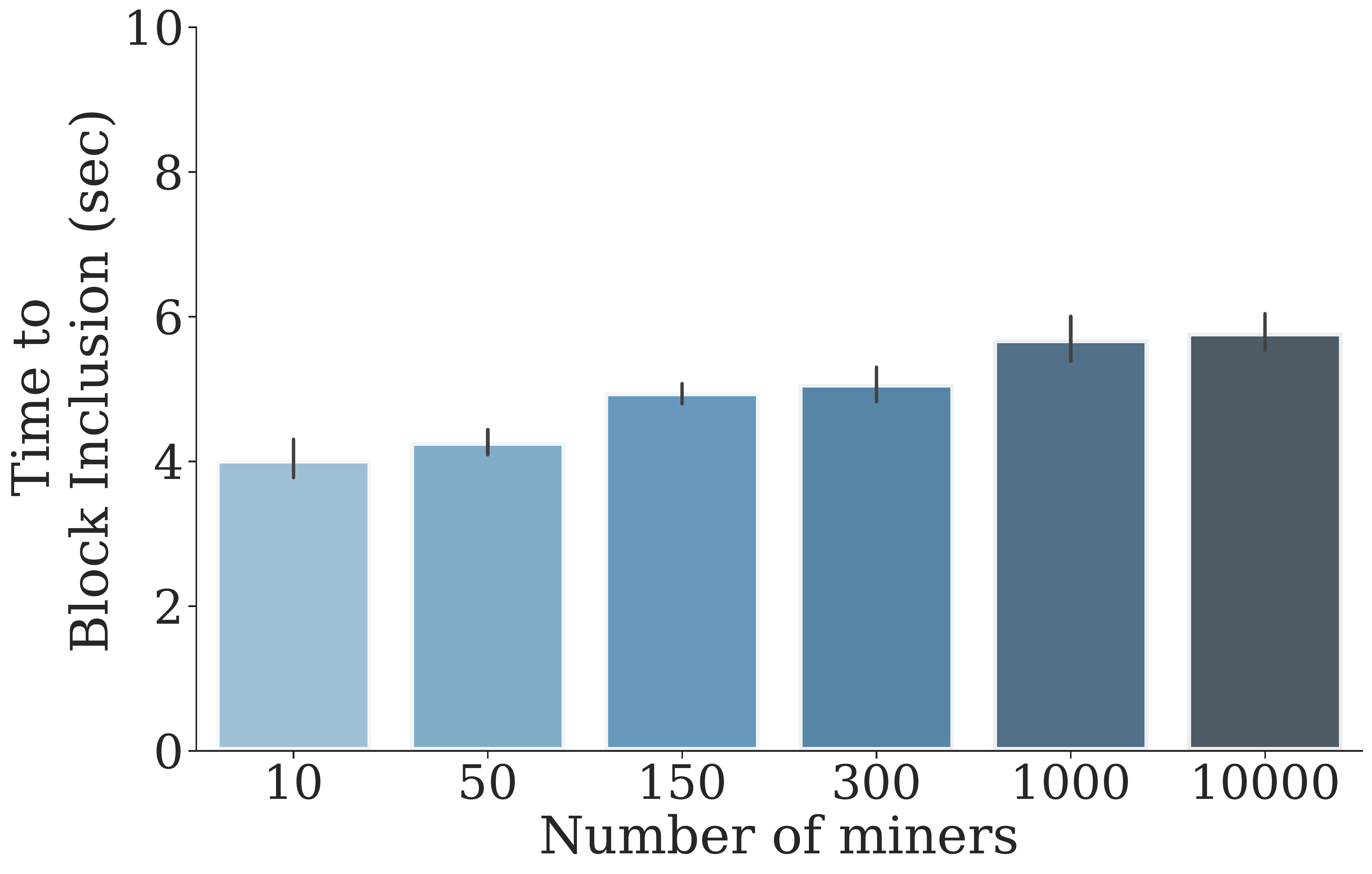} 	
	\end{subfigure}%
	\caption{(Left) Average time for a transaction to be included in a block for the vanilla algorithm used in Ethereum and for \systemname{} (i.e., 'Natural' ordering), and (Right) time for a transaction to be included in a block with \systemname{} depending on the system's size. }
	\label{fig:time_settle}
\end{figure*}

To demonstrate the effects of our new policies on block building, i.e., selecting transactions in order, we simulate a block creation process at randomly selected miners with an average block time of 12\,s, which is the block time in Ethereum.
We report the average time it takes for a transaction to be included in a block in Fig.\ref{fig:time_settle}. We compare the policy for block creation described in Section~\ref{sec:block_build} (`Natural' ordering) with the policy that is currently widely used in public blockchains, i.e., creating a block with the highest-fee transactions of the mempool (referred to as Highest Fee'). 

The average transaction latency for the 'Natural' ordering is 3 seconds, while it is around 7-8 seconds for the 'Highest Fee' strategy. Furthermore, we observe that the 'Highest Fee' strategy exhibits a wide spread along the axes, with many low-fee transactions experiencing very high latency. \systemname{}'s orders transactions according to the order with which they have been received by miners, which leads to transactions being processed sequentially and increases  fairness.  

\subsection{Protocol Overhead}

\subsubsection*{Bandwidth}

We benchmark our protocol against two baseline protocols: 'Flood' and 'PeerReview'. 'Flood' is a traditional mempool exchange protocol where miners initially send a 'Mempool' message containing a list of hashes of the transactions currently in their mempool. The receiving miner compares these hashes against its known transaction IDs and requests any missing transactions.

We also compare \systemname{} to 'PeerReview', a generic accountability protocol that could be used to monitor censorship attempts by miners~\cite{haeberlen2007peerreview}. Every miner maintains an additional log for each received message. For each miner, we assign 8 witnesses. Periodically, each miner fetches the log from the miners and checks for any injection (commission) or censorship (omission).

\begin{figure}
\centering
\includegraphics[width=1.0\linewidth]{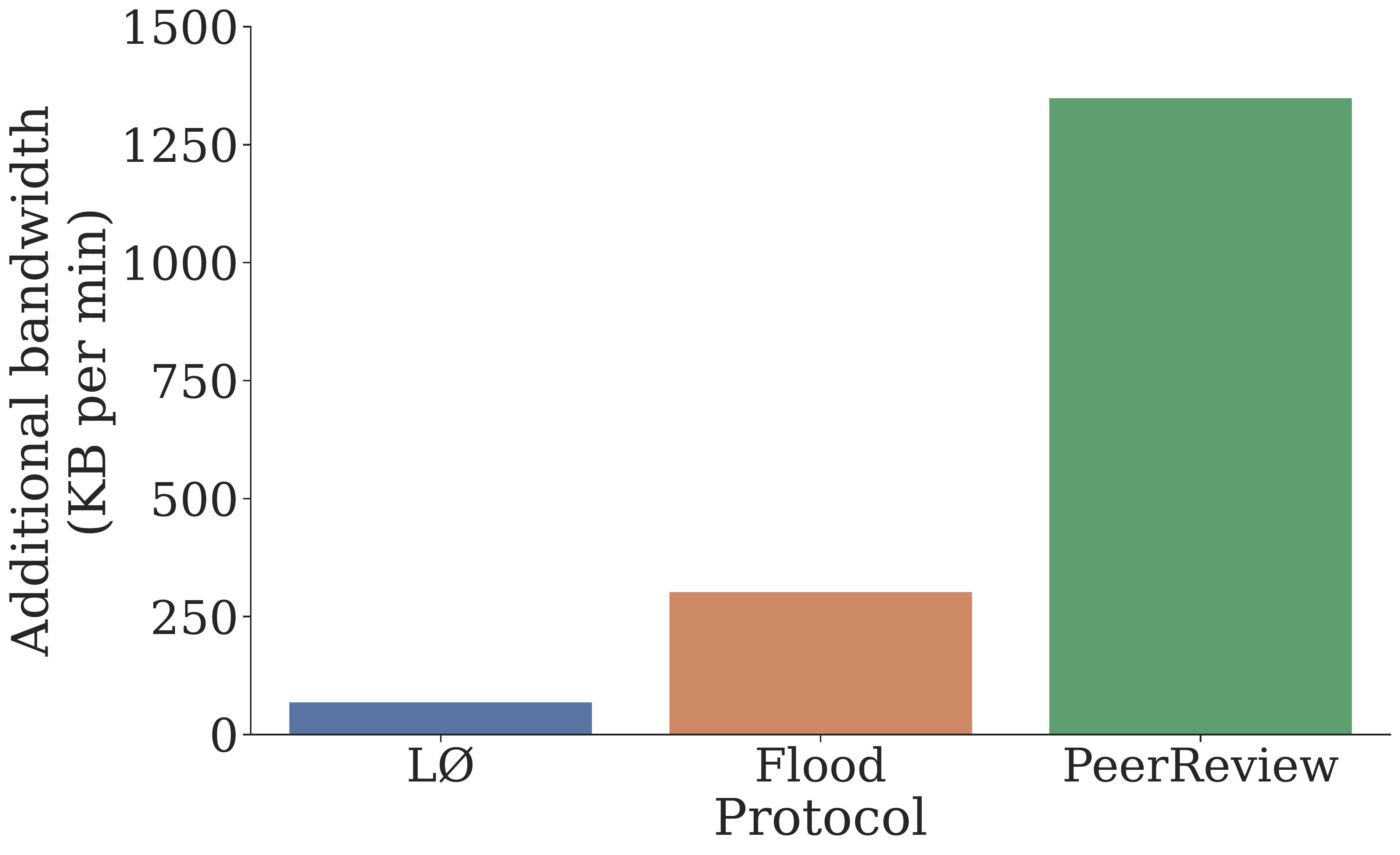}
\caption{Bandwidth overhead measured in KB per minute.}
\label{fig:bandwidth}
\end{figure}

The comparison is reported in Fig.~\ref{fig:bandwidth}. Note that we omit the bandwidth overhead for sharing transactions, as it is the same for all three protocols. Our protocol is the most bandwidth-efficient compared to the other two protocols, incurring 20 times less bandwidth overhead than PeerReview.

\subsubsection*{Memory and CPU Overhead}

\begin{figure}
\centering
\includegraphics[width=1.0\linewidth]{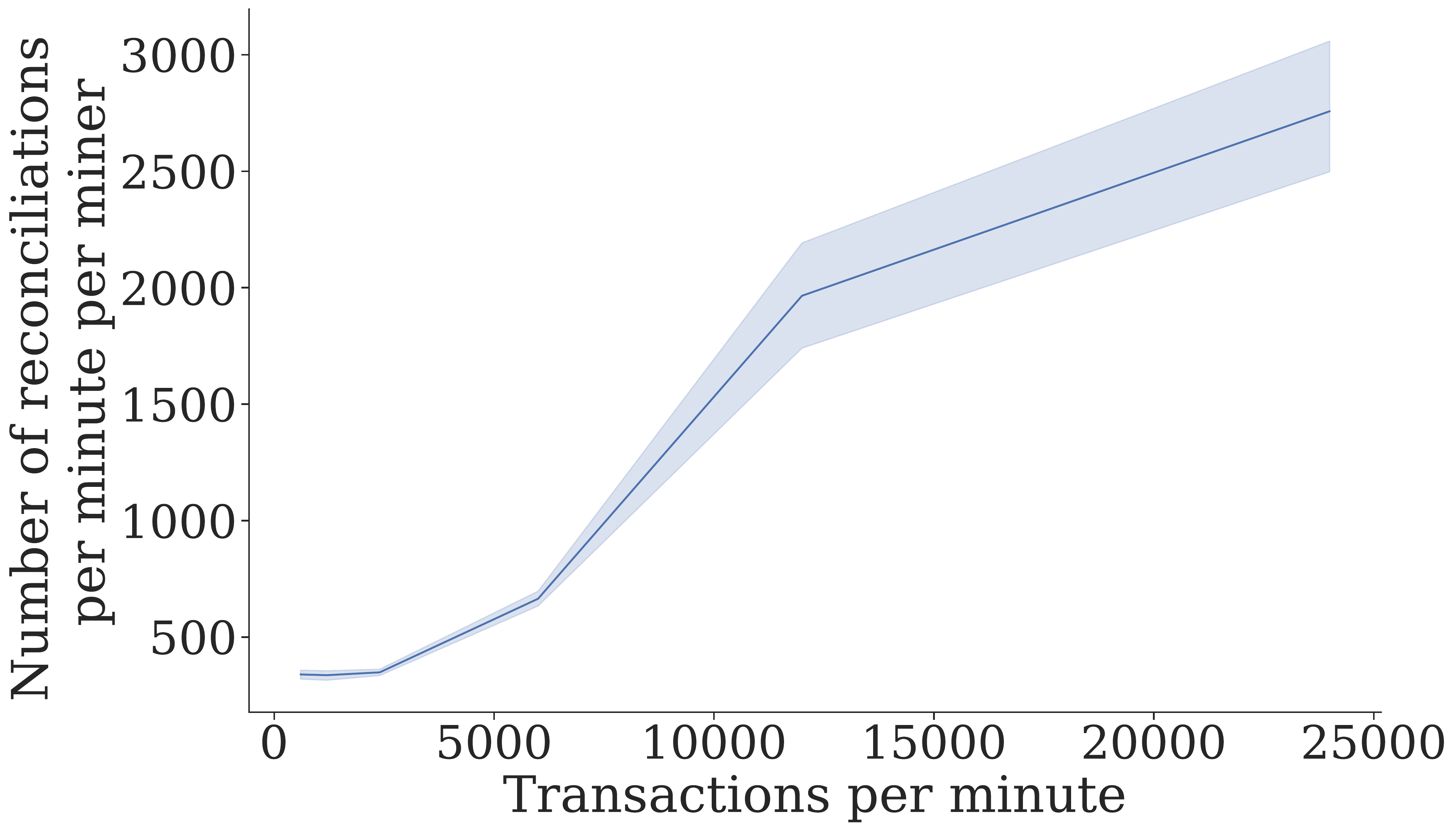}
\caption{Average number of reconciliations per minute depending on the workload (new transactions per minute).}
\label{fig:avg_recon}
\end{figure}

The overhead for encoding and decoding Minisketch scales linearly with the size of the set difference~\cite{gong2020space}. Minisketch computes a set difference with $1,000$ items in 10 seconds. To optimize the usage of the sketch, we hash-partition the mempool space into subsets, as described in~\cite{gong2020space}. Each time reconciliation fails, the node divides the mempool in half and sends an additional Minisketch for each partition. As a result of this optimization, we encode and decode all sketches required for a set difference of $1,000$ items in less than 100 ms. We report the average number of sketch reconciliations per minute per node depending on the workload in Fig.~\ref{fig:avg_recon}.

\systemname{} only requires a small additional memory overhead to store the commitments of all its neighbors. The size of the commitment depends on the workload. For example, for a workload of 120 transactions per minute, the commitment size is 1.17 KB, while for a workload of 24,000 transactions per minute, the total size of commitments can reach up to 9.36 KB. Even if the miner stores the commitments of all 10,000 nodes, it would only require 87 MB.    


\section{Related Work}
\label{sec:sota}

The problem of MEV has attracted a considerable amount of research~\cite{qin_quantifying_2022, zhou2022sok, yang2022sok}. Different MEV mitigation mechanisms can be categorized according to implementation at different layers: \emph{application, consensus, base}. 

\subsection{MEV Mitigation at the Application Layer}

Decentralized Exchanges aggregators such as Cowswap implement \textit{Batch Auctions} where orders are placed off-chain and not immediately executed, but rather, collected and aggregated to be settled in batches~\cite{cowswap}.The applications of this approach is tied to a specific application, and thus limited to specific types of MEV attacks (front running and sandwich).

A2MM is a DEX design that atomically performs optimal routing and arbitrage among the considered AMM, minimizing subsequent arbitrage transactions~\cite{zhou_a2mm_2021}. 

\subsection{MEV Mitigation at the Consensus Layer}

Proposer-builder separation (PBS) is a proposal aiming at MEV minimization~\cite{buterinPBS}.The latest iteration of this mechanism, MEV-Boost, is implemented as a middleware. It enables private communication channels between clients creating new transactions and validating nodes. However, this approach has significant trust assumptions, such as relays not reordering or censoring transactions, which empirically do not hold~\cite{yang2022sok}.

Pre-ordering solutions aim to separate transaction ordering from execution to ensure 'fair' ordering. The Helix consensus protocol~\cite{yakira_2021} guarantees random selection and ordering of transactions in blocks, relying on a randomness beacon within the consensus protocol. Aequitas~\cite{kelkar_order-fairness_2020} provides guarantees on transaction ordering within a block, but assumes a permissioned environment and introduces significant communication overhead. Pompe~\cite{zhang2020byzantine} is a Byzantine ordered consensus (BOC) protocol that outputs a transaction $t$ and a sequence number $s$ for ordering $t$. Wendy~\cite{kursawe2020wendy,kursawe2021wendy} describes ordering protocols for permissioned systems. Enforcing relative order requires building a dependency graph to prevent transactions from being included in a block before their dependencies~\cite{kelkar_order-fairness_2020,kelkar2021themis,cachin2022quick}. Enforcing fair-ordering is more resource-intensive than enforcing our accountability properties and not practical in a permissionless setting.

Heimbach and Wattenhoffer propose encrypting transaction content, ordering it, and revealing its content only after it has been ordered~\cite{heimbach_sok_2022}. This approach is implemented by Fino, which integrates MEV protection into a BFT protocol in the partial synchrony model with a DAG transport protocol~\cite{malkhi2022maximal}. Lyra~\cite{zarbafianlyra}, a Byzantine ordered consensus protocol, also uses a commit-reveal scheme and relies on Verifiable Secret Sharing (VSS). The encrypt-commit-reveal scheme is more resource-intensive than our accountable approach and requires additional trust assumptions to ensure that encrypted transactions are always revealed.

\subsection{MEV Mitigation at the Base Layer}

Secret Mempools hide the content of a transaction so that it cannot be censored, reordered, etc. 
F3B is a generic approach for online transaction encryption based on a commit-and-reveal architecture~\cite{zhang2022flash}. Ferveo is a protocol for Mempool Privacy on BFT consensus blockchains~\cite{bebel2022ferveo}. Both of these solutions assume permissioned settings. 

ZeroMEV is an existing MEV mitigation solution implemented on the base layer~\cite{ZeroMEV}. This solution is Ethereum-specific and implemented on the basis of Geth software fork as a validator execution client.
It orders transactions based on timestamps with local FIFO order. However, this solution does not provide any accountability and requires strong trust assumption as it relies on the altruism of a validator.

\section{Conclusion}
\label{sec:conclusion}

We introduced \systemname{}, an accountable base layer for permissionless blockchains. It is consensus-protocol agnostic and provides detection guarantees for various MEV attacks. \systemname{} mandates that both correct and faulty miners log all received transactions into a secure mempool data structure and exchange and record commitments on their mempool content. Any inconsistency, such as transaction withholding or equivocation, is exposed during a mempool reconciliation process with a correct miner. To ensure the exposure of faulty miners, \systemname{} simply requires correct miners to be interconnected through a network path.

We outlined the transaction manipulation attacks associated with MEV that miners might execute and mapped different attack types to the relevant stages of a transaction’s lifecycle within the protocol. Our performance evaluation demonstrates the practicality of \systemname{}. It is bandwidth and memory efficient, using only 10 MB with 10,000 miners and a workload of 20 transactions per second. Moreover, it is at least four times more bandwidth efficient than classical flooding-based mempool exchanges and processes transactions with higher fairness.


\bibliographystyle{ACM-Reference-Format}
\bibliography{references.bib}








\end{document}